\documentclass[aps,prl,twocolumn,superscriptaddress]{revtex4-1}
\usepackage{amssymb}
\usepackage{amsmath,bm}
\usepackage{graphicx,color}
\usepackage{bbm}
\usepackage[bottom]{footmisc}
\usepackage{multirow}
\usepackage[normalem]{ulem}
\usepackage{xcolor}
\usepackage{xpatch} 
\makeatletter
\usepackage{xpatch} 
\DeclareMathOperator{\Tr}{Tr}
\newcommand{\B}[1]{{\bm{#1}}}

\newcommand{\Lag}{\mathcal{L}}
\newcommand{\A}{\mathcal{A}}

\newcommand{\C}[1]{{\mathcal{#1}}}

\begin{document}
\title{Dipole Screening in Pure Shear Strain Protocols of Amorphous Solids}

\author{Chandana Mondal}
\affiliation{UGC-DAE Consortium for Scientific Research,
 Indore, 
Madhya Pradesh 452017, India}
\author{Michael Moshe} 
\affiliation{Racah Institute of Physics, The Hebrew University of Jerusalem, Jerusalem, Israel 9190}
\author{Itamar Procaccia}
\affiliation{Dept. of Chemical Physics, The Weizmann Institute of Science, Rehovot 76100, Israel}
\author{Saikat Roy}
\affiliation{Department of Chemical Engineering, Indian Institute of Technology Ropar, Punjab 140001, India}

\begin{abstract}
When amorphous solids are subjected to simple or pure strain, they exhibit elastic increase in
stress, punctuated by plastic events that become denser (in strain) upon increasing the system size. It is customary to assume in theoretical models that the stress released in each plastic event is redistributed according to the linear Eshelby kernel, causing avalanches of additional stress release. Here we
demonstrate that contrary to the uniform affine strain resulting from simple or pure strain, each
plastic event is associated with a non-uniform strain that gives rise to a displacement field that contains quadrupolar and dipolar charges that typically screen the linear elastic phenomenology and introduce anomalous length-scales and influence the form of the stress redistribution. An important
question that opens up is how to take this into account in elasto-plastic models of shear induced
phenomena like shear-banding.
\end{abstract}
\maketitle
{\bf Introduction:} Amorphous solids, including a host of substances, from metallic and silica glasses to gels and powders, pose exciting theoretical challenges in understanding their mechanical properties and failure modes \cite{99ML,06ML}. Contrary to perfect elastic media, amorphous solids experience plastic events in response to any amount of external stress \cite{10KLP,11HKLP}. For large external shear strain, accumulation of plastic responses can lead to mechanical failure of amorphous solids through shear-banding and the appearance of cracks \cite{12DHP,13DHP}. 

The phenomenon of shear banding is a limiting factor for the usefulness of amorphous solids in applications, and as such it attracted enormous amount of attention, especially in the context of failure under pure or simple shear. Both simulations and experiments abound, leading to an active developments of models which are collectively known as ‘elastoplastic’ models \cite{98HL,06Sol,17NFMB}. While the available models differ in detail, elastoplastic models handle the material as a collection of ‘mesoscopic’ blocks alternating between an elastic behavior and plastic relaxation, when they are loaded above a threshold. Plastic relaxation events redistribute stresses in the system; the lost stress is distributed between all the other cells, such that the amount of stress that each cell receives is determined by the `Eshelby kernel', a function that was computed by Eshelby in the 1950’s for a quadrupolar strain perturbation in a perfectly elastic medium \cite{57Esh}. This protocol can induce avalanches of ‘plastic events’ and at a certain global strain the avalanche causes a shear band.

Even before the onset of shear banding, plastic responses can not only renormalize the elastic properties of the system, but can also induce a qualitative deviation from an elastic response. This puts doubts on the relevance of Eshelby’s kernel as solved within linear elasticity theory. In fact, we have recently developed a geometric model of mechanical screening via quadrupole and dipole elastic charges, which predicted new phenomenology within linear response, that was later fully observed in experimental and numerical systems \cite{21LMMPRS,22MMPRSZ,22BMP,22KMPS,22CMP}. In this theory the response to local perturbation is screened by various geometric multipoles.

It therefore behooves on us to examine the role of screening before the onset of shear banding, an issue which appears fundamental to elastoplastic models in general. If dipole screening is non-existent at small strains, then the common protocol of using the classical Eshelby’s kernel is justified. If, however, dipole screening exists at small strains, it suggests that a modified version of the classical Eshelby kernel should be developed. The aim of this Letter is to test the screening mode prior to shear banding. We provide theoretical and simulational evidence below that in fact every plastic event creates quadrupolar and dipolar effective charges in the displacement field that follows the event. We demonstrate these issues in the context of pure shear strain of a generic model of amorphous solids, but elastoplastic modeling of simple strain will suffer from the very similar issues. 

{\bf Simulations}: To demonstrate the issues we chose as our example frictional granular matter, to be as close as possible to realizable experiments. Our simulations employed amorphous granular assemblies of $16000$ disks, half of which have a radius $R_1=0.35$ and the other half with $R_2=0.49$. The details of the contact forces and the protocols for creating an equilibrated configuration at any desired pressure $P_0$ are standard, and are presented in the appendix.

\begin{figure}
	\includegraphics[width=0.9\linewidth]{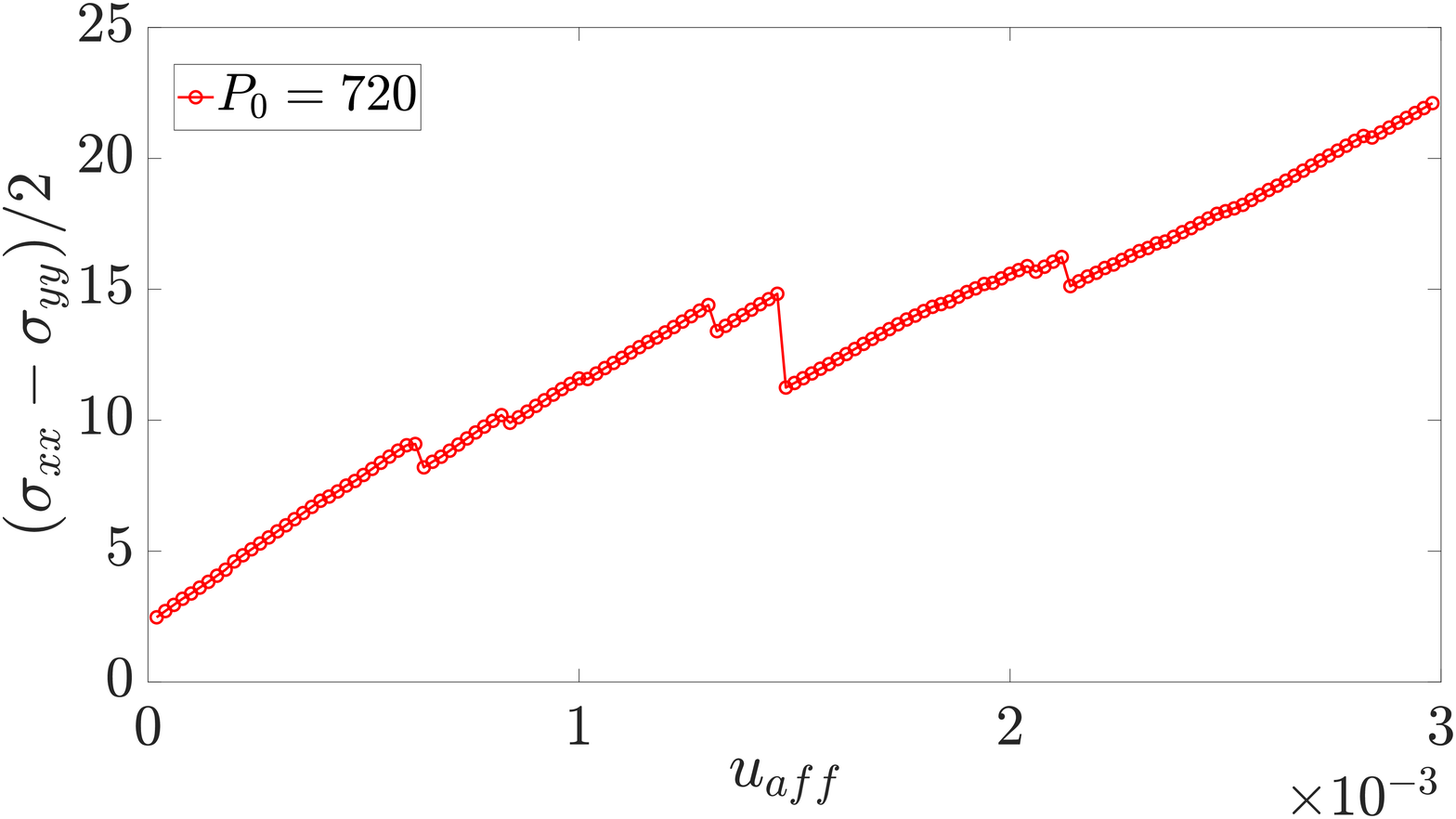}
	\includegraphics[width=0.9\linewidth]{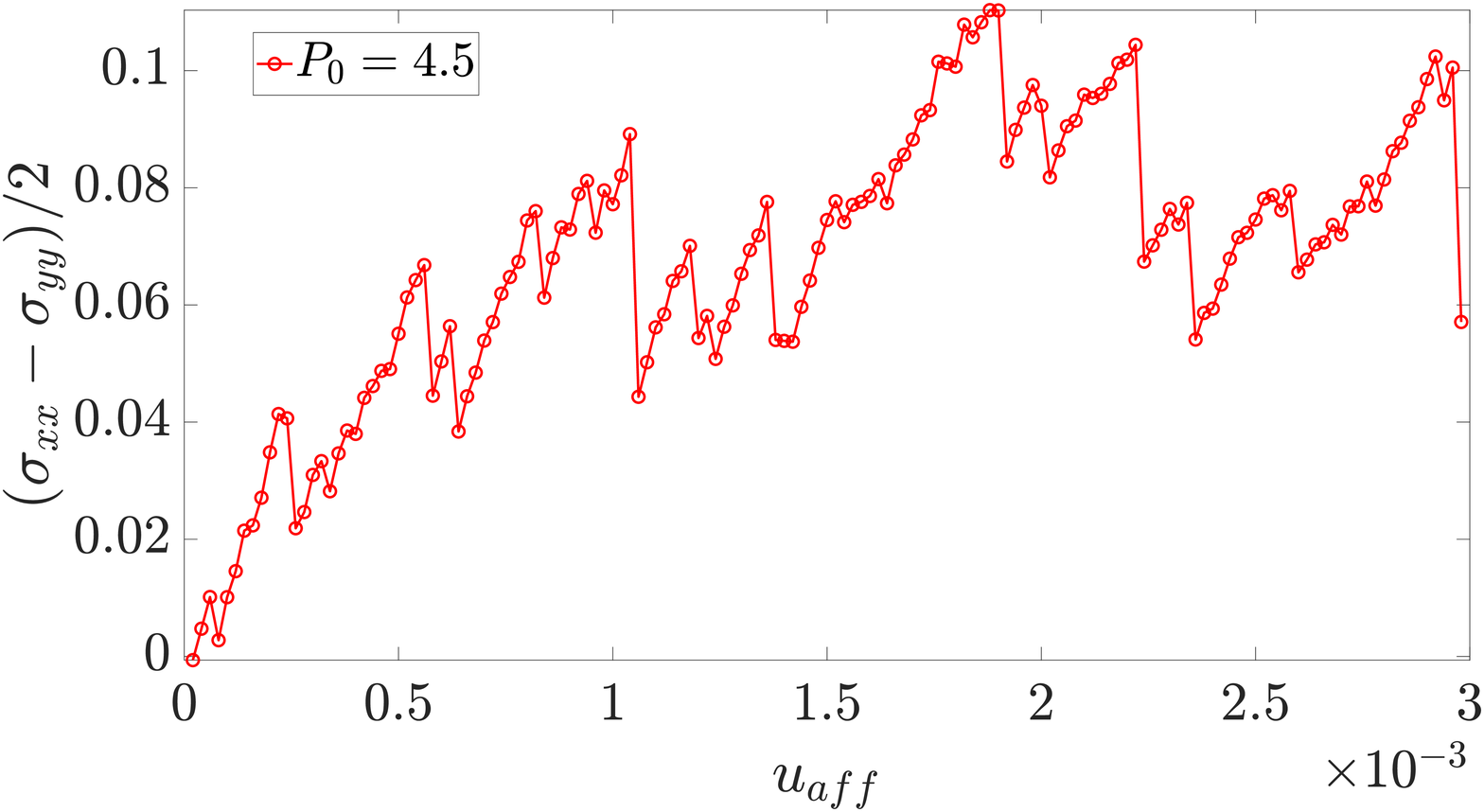}
	\caption{Shear stress vs accumulated affine strain in pure shear. Shown are two initial pressures $P_0=720$ (upper panel)
	and $P_0=4.5$, our highest and lowest pressures. In both cases one sees elastic increase in stress punctuated by plastic events, that are denser and more violent when the pressure is smaller. }
	\label{shear}
	\end{figure}

Having a mechanically stable configurations at different  pressure values $P_0$ with box dimensions $Lx_0$ and $Ly_0$ along x and y directions respectively, we apply volume-preserving pure shear on the samples, involving the following steps: (i) we reduce the box lengths along $x$ by $0.00002\%$ and expand it along $y$ directions such that volume of the system remains constant at $Lx_0\times Ly_0$; (ii) we run constant NVE simulation, until the force and torque on each and every particle are smaller than $10^{-7}$ in reduced units. We repeat these two steps 2000 times for all the pressures. We measure the instantaneous pressure $P$ and the accumulated {\em affine} strain 
 \begin{equation}
u_{\rm aff}\equiv \frac{1}{2}\big(\frac{Lx_0-Lx}{Lx_0}+\frac{Ly-Ly_0}{Ly_0}\big) \ ,
\end{equation} 
where Lx and Ly are the instantaneous box-lengths along x and y directions respectively. Typical shear stress vs. (affine) strain plots are shown in Fig.~\ref{shear} for our lowest and highest initial pressures. As is usual in such simulations, we observe intervals of increase in stress when the strain increases, interrupted by sharp drops in stress due to plastic events. These are the events that we focus on next. 

{\bf Displacement fields associated with plasticity}: presently we focus on the displacement field that is triggered by the plastic drop. Denoting the positions of our $N$ disks before and after the event as $\B r_i^a$ and $\B r_i^b$ respectively, we compute  the displacement field as $\B d_i\equiv \B r_i^a-\B r_i^b$. Next we compute the total strain field as
\begin{equation}
	u_{ij} = 0.5(\nabla_i d_j  + \nabla_j  d_i )
	\label{defu}
\end{equation}
The non-affine strain $\B u_q$ is obtained by subtracting the affine strain  generated in the last step from $u_{\rm tot}$, 
\begin{eqnarray}
&&u^q_{11}\equiv u_{11} - \frac{1}{2}\big(\frac{Lx^b-Lx^a}{Lx^b}\Big)\ ,\nonumber\\
&&u^q_{22}\equiv u_{22}-\frac{1}{2}\big(\frac{Ly^a-Ly^b}{Ly^b}\big) \ , \nonumber\\
&&u^q_{12}\equiv u_{12}\ , \quad u^q_{21}\equiv u_{21} \ .
\end{eqnarray}
where again `a' and `b' refer to after and before. Having the non-affine strain we decompose it into its trace and its traceless components (cf. Ref.~\cite{15MSK} page 6):
\begin{equation}
\B	u^q = m\B I +Q \B u^{ts} \ ,
\end{equation}
where $\B I$ is the identity tensor and $\B u^{ts}$ a traceless symmetric tensor. In the last equation $m=0.5 \Tr u_q$ and 
\begin{equation}
Q^2 =  (u^{ts}_{11})^2  +  (u^{ts}_{22})^2 \ .
\label{defQ}
\end{equation}
The quadrupolar charge $Q$ is obtained as the square root, and its orientation is computed from \cite{15MSK}:
\begin{equation}
	\Theta = 0.5 \arctan ((u^{ts}_{12})/(u^{ts}_{11}) ) \ .
	\label{deftheta}
	\end{equation}

\begin{figure}
	\includegraphics[width=0.9\linewidth]{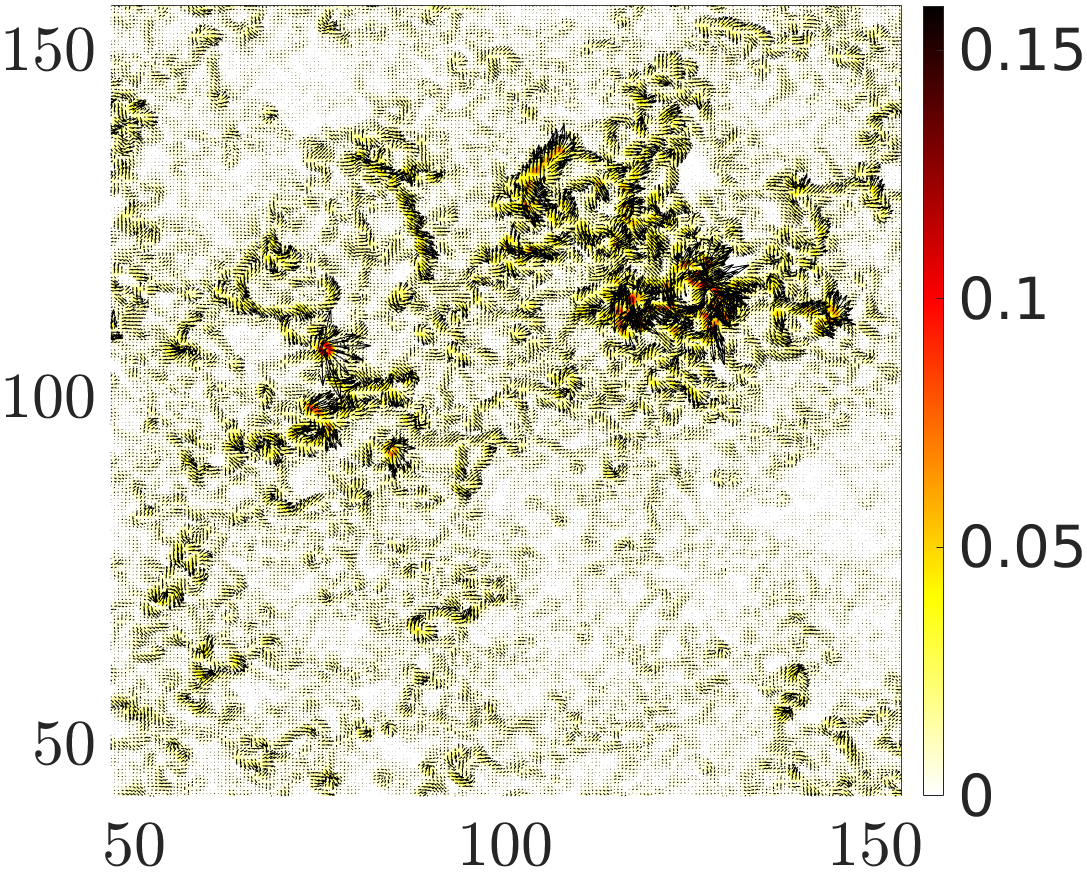}
	\includegraphics[width=0.9\linewidth]{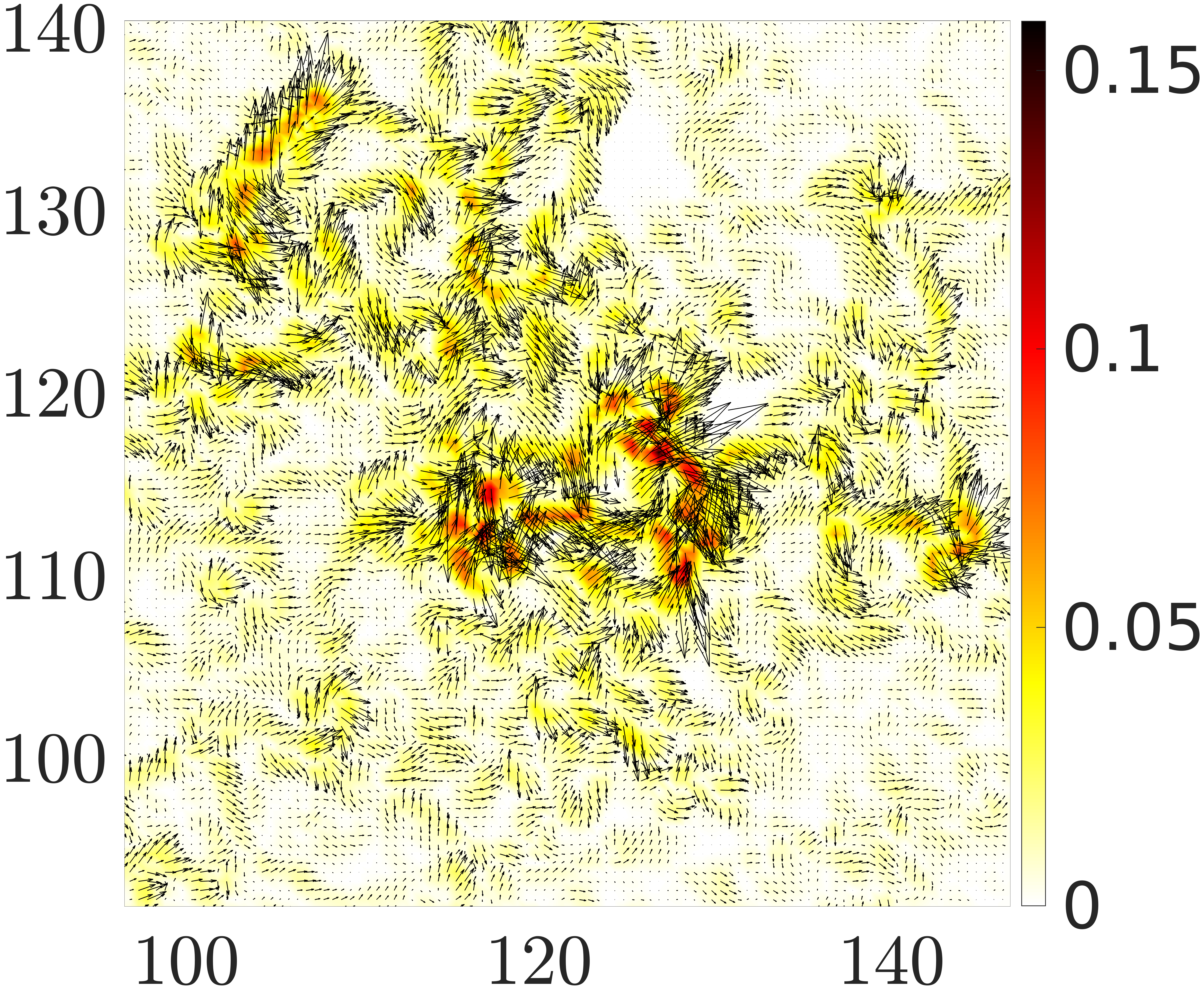}
	\caption{Heat map of the quadrupolar field for our system after a plastic event at a lower pressure
		$P_0=4.5$. The darker region indicate high values of $Q$ cf. Eq.~(\ref{defQ}), and light region low values. The arrows are in the direction of the angle $\Theta$, cf. Eq.~(\ref{deftheta}). In the upper panel we show the whole system and then a zoom into the most active region.  }
	\label{Q}
\end{figure}

A typical map of the quadrupolar fields computed in this fashion, with the arrows in the direction of the angle $\Theta$, are shown in Fig.~\ref{Q} for the low pressure exhibited in Fig.~\ref{shear}. 
The upper panel shows the map for the whole system and below a zoom on the most active region. The map for the high pressure is similar, but with a difference in scale - the quadrupolar field is considerably more intense in the case of lower pressure. The arrows are pointing in the direction of the angle $\Theta$, note that here there is no preferred angle with respect to the principal stress axis \cite{12DHP,13DHP}.

\begin{figure}
	\includegraphics[width=0.9\linewidth]{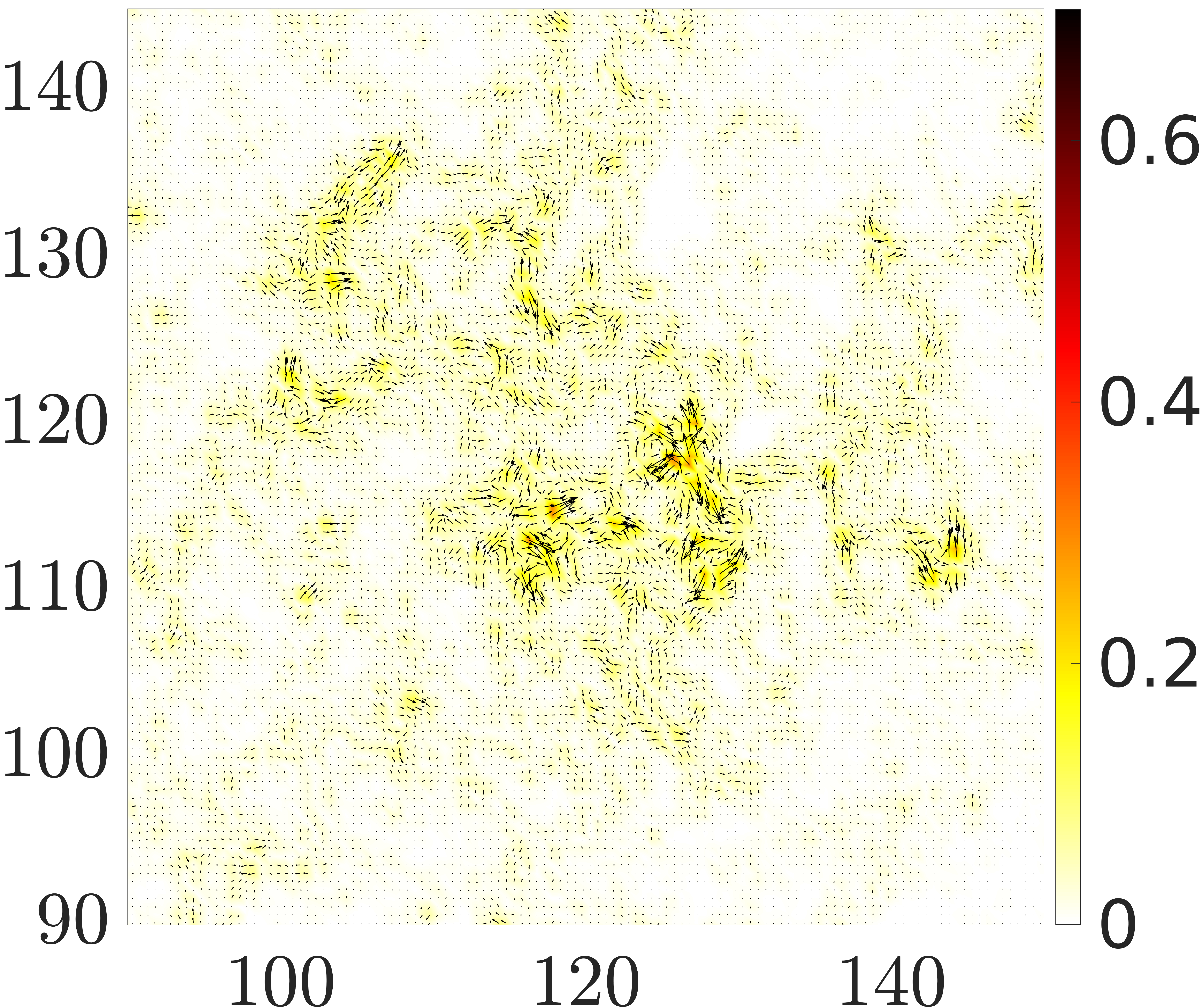}
	\vskip 0.3 cm
	\includegraphics[width=0.9\linewidth]{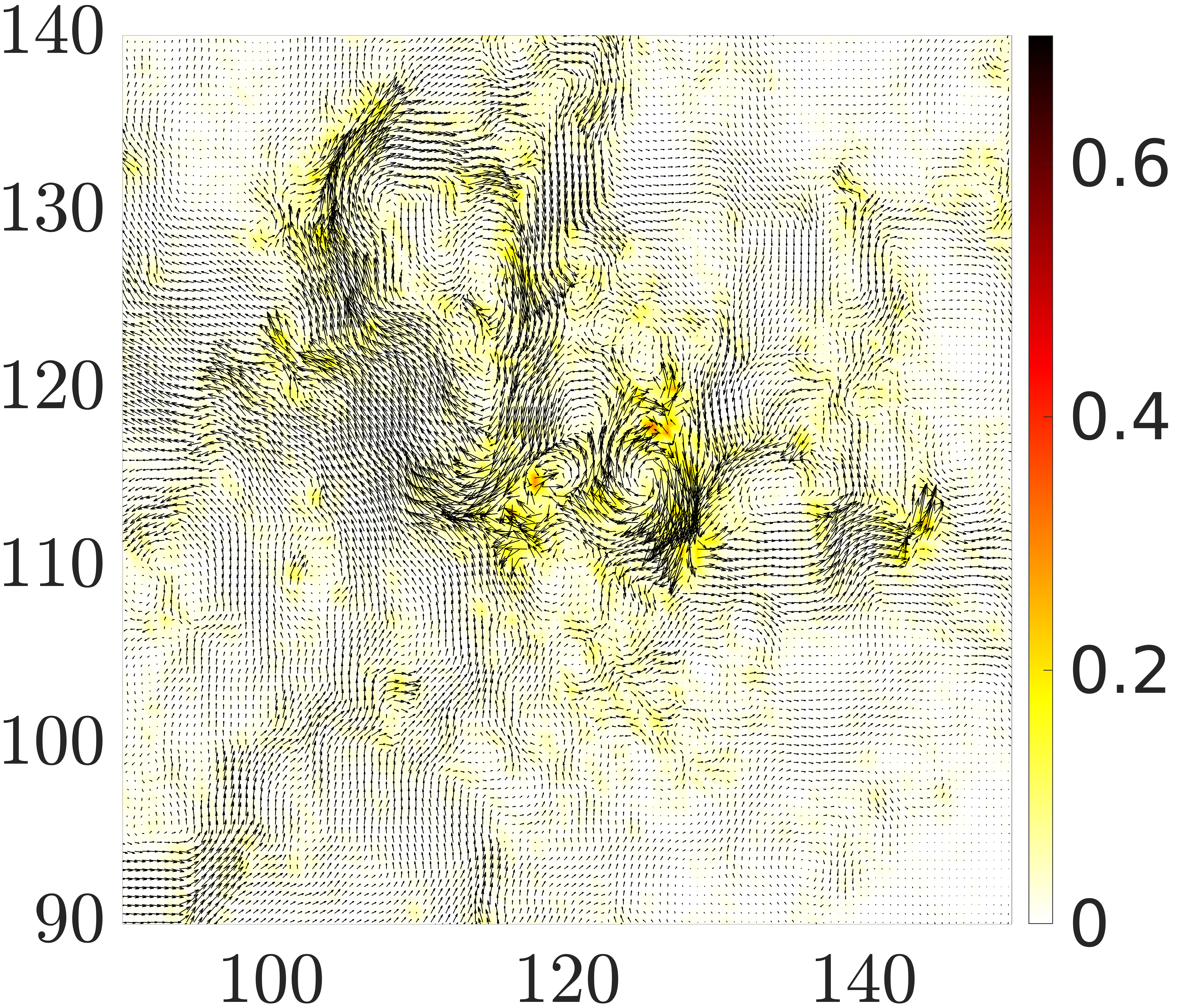}
	\caption{Upper panel: heat map of the dipole field  $\C P^\alpha\equiv \partial_\beta Q^{\alpha\beta}$ for $P_0=4.5$, in the window of the lower panel of Fig.~\ref{Q}. Lower panel: minus the displacement field in the same window. The arrows in both panels are in the local direction of the respective field.}
	\label{dip}
\end{figure}

Since the quadrupolar field is obviously non-uniform, we expect that its divergence would be quite important. Thus we swiftly proceed to compute the dipolar field $\B {\C P}$, as the latter is expected to be crucial for the way stress is distributed as a result of the plastic event. The dipolar field is simply computed as $\C P^\alpha\equiv \partial_\beta Q^{\alpha\beta}$ \cite{21LMMPRS,22MMPRSZ,22BMP,22KMPS,22CMP}.
In the upper panel of Fig.~\ref{dip} we present the divergence of the quadruopolar field $\B Q$ that is shown in lower panel of Fig.~\ref{Q}. At this point the important observation is that this field is not zero.  

{\bf Theoretical considerations}: examining the dipolar heat maps and the direction of the dipoles one gets the impression that this field is quite disordered, with arrows pointing in all directions. 
In fact, the theory presented in Refs.~\cite{21LMMPRS,22MMPRSZ,22BMP,22KMPS,22CMP} predicts that the dipole field should be proportional to the displacement field, and the latter is indeed quite disordered. As a brief summary of the theory, we recall that classical elasticity in two dimensions can be derived from a Lagrangian 
by minimizing the energy $F$, 
\begin{eqnarray}
	F &=& \int \Lag \, \mathrm{d} x \mathrm{d} y- \oint t^\beta  d_\beta \, \mathrm{d}S \ , \nonumber\\
	\C L &=&  \frac{1}{2} A^{\alpha\beta\gamma\delta} u_{\alpha\beta}u_{\gamma\delta} =\frac{1}{2}\sigma^{\alpha\beta} u_{\alpha\beta} \ ,
	\label{basic}
\end{eqnarray}
where $\B A$ is the usual elastic tensor, and $\mathrm{d} S$ is the area element on the boundary. Minimizing the energy one derives the classical result $\partial_\alpha \sigma^{\alpha\beta} = 0.$ In Refs.~\cite{21LMMPRS,22MMPRSZ,22BMP,22KMPS,22CMP} it was shown that in the presence of quadrupolar plastic response the elastic tensor is renormalized, yielding a new tensor $ \tilde A^{\alpha\beta\gamma\delta}$ and a renormalized stress field satisfying yet the same equation $\partial_\alpha \tilde \sigma^{\alpha\beta} = 0.$ On the other hand, once there exist gradients of the quadrupolar field, generating dipoles, $\C P^\alpha\equiv \partial_\beta Q^{\alpha\beta}$, the appropriate Lagrangian takes into account the dipoles in the form 
\begin{equation}
	\begin{split}
		\Lag &=  \frac{1}{2} \tilde{\A}^{\mu\nu\rho\sigma} u_{\mu\nu}u_{\rho\sigma} + 
		\frac{1}{2} \Lambda_{\alpha\beta} \partial_\mu Q^{\mu\alpha}  \partial_\nu Q^{\nu\beta}
		+ \Gamma_{\alpha}^{\,\,\beta} \partial_\mu Q^{\mu\alpha} d_{\beta} \ ,
	\end{split}
	\label{lagdip}
\end{equation} 
where the tensors $\B \Lambda$ and $\B \Gamma$ are new coupling tensors that do not exist in classical elasticity theory. Minimizing the energy associated with this Lagrangian results in a new equation satisfied by the stress field, 
\begin{equation}
	\partial_\alpha \sigma^{\alpha\beta} =- \Gamma_{\alpha}^{\beta} \C P^{\alpha} \ .
	\label{eq:Equilibrium2}
\end{equation}
One should note that this equation breaks translational symmetry as explained in \cite{21LMMPRS,22MMPRSZ,22BMP,22KMPS,22CMP}. 
In isotropic homogeneous media the coupling tensors simplify, reading 
$\Gamma_{\alpha}^{\beta} = \mu_1 g^\alpha_\beta$, 
$\Lambda^{\alpha\beta} = \mu_2 g^{\alpha\beta}$ where $\B g$ is the Euclidean metric tensor, and
$\mu_1,\mu_2$ being scalar novel moduli that do not exist in classical elasticity. Finally, and importantly for our purposes here, it was shown that the diploar field satisfies an equation
\begin{equation}
	\B {\C P} =-\kappa^2 \B d\ , 
	\label{Pvsd}
\end{equation}
 where $\kappa$ is an inverse scale that acts as a screening parameter. This is the reason that the dipole field appears as chaotic as the displacement field. 
 To establish that the theory is relevant in the present context we test Eq.~(\ref{Pvsd}) in our simulations.

\begin{figure}
	\includegraphics[width=0.9\linewidth]{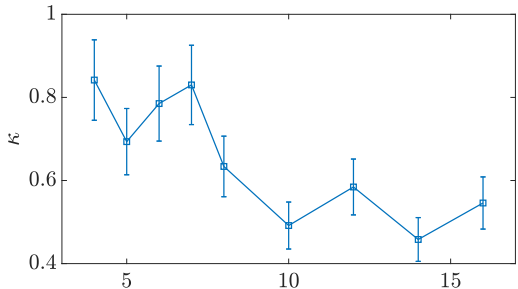}
	\caption{The screening parameter $\kappa\approx 0.68\pm 0.2$ computed by dividing the two integrals
		in Eq.~(\ref{dipole1}) computed on square loops of different sizes and taking the square root. Results pertain to an average over 20 central grid points, error bars reflect statistical error.}
	\label{ratio}
\end{figure}

{\bf Test of theory:} Equation (\ref{Pvsd}) is an important constitutive relation that is predicted by the theory, but was never put to a direct test as we can do here. In the lower panel of Fig.~\ref{dip} we show (minus) the displacement field from which the data of the upper panel of Fig.~\ref{dip} was computed, following the recipe presented above. Indeed, to the eye it appears that the two fields are proportional to each other, as expected from the theory. To provide a quantitative test we can integrate Eq.~(\ref{Pvsd}) around any closed loop and test whether 
\begin{equation}
	\oint _{\partial \Omega }\B{\C P}(x,y)\cdot \mathbf{n}	\, \text{dl}=-\kappa^2	\oint _{\partial \Omega }\B d(x,y)\cdot \mathbf{n}	\, \text{dl} \ ,
	\label{dipole1}
\end{equation} 
where $\B n$ is the unit vector normal to the integration path, pointing outward. In the present case it is natural to choose square trajectories for the integrals, thus using the $x$ component of the field for paths along $y$ and the $y$ components for paths along $x$, with appropriate signs. We have chosen 20 central points on the grid that was used to digitize the displacement field, and for each such point we computed the two line integrals on squares of edge sizes 6-23. After taking the ratio of the two integrals in Eq.~(\ref{dipole1}) we computed the square root and averaged $\kappa$ over the twenty central points. One should point out that the protocol described in Eqs.~(\ref{defu})-(\ref{deftheta}), including the computation of the divergence of the quadrupolar field at the end, is not free of numerical noise (at each step). It is therefore quite remarkable that the resulting value of $\kappa$ as shown in Fig.~\ref{ratio} is quite stable, $\kappa\approx 0.68\pm 0.2$. A priori it is not even guaranteed that the ratio of the two integrals would be negative definite, resulting in a real value of $\kappa$. We thus interpret the results of the calculation as a strong 
support for the constitutive relation Eq.~(\ref{Pvsd}). 

Having demonstrated that generic plastic drops induce a displacement field that is typically exhibiting effective dipoles, we must realize that the fundamental change in physics that is embodied in Eq.~(\ref{eq:Equilibrium2}) requires reassessment of the redistribution of the stress that is lost in the plastic drop. It is no longer likely that the regular power law decay of the Eshelby kernel would describe properly this redistribution. It was amply demonstrated that the appearance of dipoles results in the introduction of a typical scale (which is actually of the order of $\kappa^{-1}$) and it can even reverse the displacement field that is expected from linear elasticity to decay monotonically. It is our proposition, on the basis of the analysis presented above, that the consequences of these results in the context of elastoplastic models should be carefully assessed. 

In the future it would be important to seek similar clarification of the role of dipole charges also in three spatial dimensions. Contrary to the Hexatic \cite{78HN} and the Kosterlitz-Thouless \cite{16Kos} phase transitions which are relevant in two-dimensions, the presence of dipoles as divergences of quadrupolar fields has been recently demonstrated in three dimensions \cite{22CMP}. The use of Eshelby kernels that were derived for purely elastic media must be reassessed. 
	
\vskip 0.5 cm

{\bf Appendix}

\vskip 0.3 cm
	
The contact forces, which include both normal and tangential components due to friction, are modeled according to the discrete element method developed by Cundall and Strack \cite{79CS}, combining a Hertzian normal force and a tangential Mindlin component. Full details of these forces and the equations of motion solved can be found in Refs.~\cite{01SEGHLP,19CGPP,19CGPPa,21LMPR}. Simulations are performed using the open source codes, LAMMPS \cite{95Pli} and LIGGGHTS \cite{12KGHAP} to properly keep track of both the normal and the history-dependent tangential force. Initially, the grains are placed randomly in a large two dimensional box while forbidding the existence of overlaps or contacts. The system is then isotropically compressed along $x$ and $y$ directions while integrating Newton's second law with total forces and (scalar) torques acting on particle $i$ given by
$\B F_{i}= \sum_{j}\B F^{(n)}_{ij} + \B F^{(t)}_{ij}$, and
$\tau_{i}= \sum_{j}\tau_{ij}$
with
\begin{equation}
	\tau_{ij}\equiv-\frac{1}{2}\left({\B r}_{ij} \times\B F^{(t)}_{ij}\right)\cdot{\B e}_z
\end{equation}
the torque exerted by $j$ onto $i$.
Compression is performed using a series of steps which involve: (i) one MD step during which we reduce the box lengths along $x$ and $y$ directions by $0.002\%$; (ii) a constant NVE run, until the force and torque on each and every particle are smaller than $10^{-7}$ in reduced units. This guarantees that the cell remains square throughout the process. We repeat these compression and relaxation cycles until the system attains a jammed (mechanically balanced) configuration at the different final pressure, fixed to $P_0= 4.5, 18, 72.0, 144, 288, 720$ (in reduced units)~\cite{21LMPR}.
Of course, in the final \emph{mechanically equilibrated states} obtained at the end of compression the total forces and torques $\B F_i$ and $\tau_i$ vanish with $10^{-7}$ accuracy, as well as all the velocities.

{\bf Acknowledgments}: This work has been supported in part by the the joint grant between the Israel Science Foundation and the National Science Foundation of China, and by the Minerva Foundation, Munich, Germany.

\bibliography{ALL.anomalous}

\end{document}